# Transforming heat transfer with thermal metamaterials and devices

*Ying Li[1,2,3], Wei Li[4], Tiancheng Han[5], Xu Zheng[6], Jiaxin Li[7,1], Baowen Li[8,6*], Shanhui Fan[4*], and Cheng-Wei Qiu[1*]*

[1]Department of Electrical and Computer Engineering, National University of Singapore, Singapore 117583, Singapore

[2]Interdisciplinary Center for Quantum Information, State Key Laboratory of Modern Optical Instrumentation, College of Information Science and Electronic Engineering, Zhejiang University, Hangzhou 310027, China

[3]ZJU-Hangzhou Global Science and Technology Innovation Center, Key Lab. of Advanced Micro/Nano Electronic Devices & Smart Systems of Zhejiang, Zhejiang University, Hangzhou 310027, China

[4]Department of Electrical Engineering, Ginzton Laboratory, Stanford University, Stanford, California 94305, USA

[5]National Engineering Research Center of Electromagnetic Radiation Control Materials, State Key Laboratory of Electronic Thin Film and Integrated Devices, University of Electronic Science and Technology of China, Chengdu 610054, China

[6]Department of Physics, University of Colorado, Boulder, CO 80309, USA

[7]School of Mechatronics Engineering, Harbin Institute of Technology, Harbin 150001, China

[8]Paul M. Rady Department of Mechanical Engineering, University of Colorado, Boulder, CO 80309, USA

*e-mail: baowen.Li@colorado.edu, shanhui@stanford.edu, chengwei.qiu@nus.edu.sg

**Abstract |** The demand for sophisticated tools and approaches in heat management and control has triggered fast development of emerging fields including conductive thermal metamaterials, nanophononics, far-field and near-field radiative thermal management, etc. In this review, we cast a unified perspective on the control of heat transfer, based on which the related studies can be considered as complementary paradigms toward manipulating physical parameters and realizing unprecedented phenomena in heat transfer using artificial structures, such as thermal conductivity in heat conduction, thermal emissivity in radiation, and properties related to multi-physical effects. The review is divided into three parts that focus on the three main categories of heat flow control, respectively. Thermal conduction and radiation are emphasized in the first and second parts at both macro- and micro-scale. The third part discusses the efforts to actively introduce heat sources or tune the material parameters with multi-physical effects in both conduction and radiation, including works using thermal convection. We conclude the review with challenges in this research topic and new possibilities about topological thermal effects, heat waves, and quantum thermal effects.

**Main text**

### I. Introduction

Heat transfer is a fundamental phenomenon of energy transport[1], generally induced by a temperature difference in space. The major concerns of heat transfer are temperature and heat flux management/control: heating / cooling targets (*e.g.* human body) to suitable temperatures and energy harvesting: converting the thermal energy from a heat source (*e.g.* the sun) to work or other forms of energy. Today, controlling heat is becoming unprecedentedly important and challenging to meet the critical problems of global warming, energy crisis, and the heating of electronic devices, which require advanced tools to manipulate heat transfer in various forms at different length scales.

In recent years, the developments on material science and physics have stimulated a renaissance of research on heat transfer. On one hand, new approaches are emerging for traditional purposes of heating/cooling and energy harvesting, in terms of their better efficiency, accuracy, adaptiveness, tunability, compactivity, etc. On the other hand, new applications have been proposed to treat heat as an information carrier, and to manipulate it for communication, detection/anti-detection, and calculation. These attempts have generated several active research fields so far such as conductive thermal metamaterials[2,3], nanophononics[4], far-field[5,6] and near-field[7] thermal radiation management, etc. In this review, we cover the related studies on the various thermal metamaterials and devices in a unified theme of manipulating heat transfer through their unusual material properties such as thermal conductivity and emissivity, corresponding to the two major forms of heat transfer: conduction and radiation. In addition, active approaches using material properties related to multi-physical effects are also included.

Heat conduction is the major form of heat transport in solids. Engineering the thermal conductivity is a central issue of its manipulation. At macroscale, such idea has been demonstrated in various conductive thermal metamaterials. The first proposal is a thermal cloak[8,9] theoretically designed to render an object undetectable through temperature measurements. It was then experimentally realized with layered structures to achieve the required anisotropic thermal conductivity[10]. At microscale, the design of thermal materials and devices is mostly based on the mechanism of phonon transport and scattering. Many works in the field of nanophononics[11] proposed to engineer the thermal conductivity with coherent phonon transport[12], phononic crystals[13], and local resonances[14]. In particular, nonlinear phononics have made the thermal diode and thermal transistor possible, and even thermal logic gates and thermal memory have been demonstrated[4].

Thermal radiation carries heat in the form of photons or electromagnetic waves. When the distance between the objects is larger than the wavelength of thermal radiation (thermal wavelength), the radiation is in the far-field regime. Far-field thermal radiation from high-temperature objects plays vital roles in solar energy harvesting, incandescent lighting, radiative cooling, etc. Around room temperature, far-field radiation is mainly in the infrared, which has become an important source for detection such as night vision. Recent advances in manipulating far-field thermal radiation using sub-wavelength structures[15,16] have offered possibilities that are drastically different from the conventional cases governed by the Plank's law and Kirchoff's law, with promising applications on energy and information processing. When the distance between the objects is sub-wavelength, the radiation is in the near-field regime[17], which can be drastically enhanced beyond the far-field case[18], indicating novel applications at microscale.

Above research fields are mostly focused on developing passive approaches to control heat transfer. Besides, heat transfer can be actively manipulated with multi-physical effects. The most common heat source/drain is based on Joule heating and Peltier cooling. Introducing thermoelectric components with feedback enables great control over heat conduction, including apparent negative thermal conductivity, at the cost of energy input[19]. Another important application of multi-physical effects is the tailoring or tuning of material parameters like conductivity[20] or emissivity[21] with other physical quantities such as electric fields. Heat convection can be considered as a multi-physical effect as well, because it is coupled with the velocity field, which is a useful tool to manipulate heat transfer[22].

The review is concluded with challenges and possible new directions. Since heat transfer is a very broad topic, we choose to focus on works targeted at controlling it using artificial structures and devices like metamaterials, nanophononic crystals, and systems with active sources. Therefore, we avoid too much discussion on works that study the phenomenon, mechanism, or certain natural/synthesized materials. Studies that aim at using heat transfer for other purposes (heat engine, thermal memory, etc.) are beyond our scope. Traditional techniques and devices like heat sink, heat pipe, etc. are also not discussed here.

## II. Manipulating heat conduction

**Macroscopic approaches**

Today, many active researches on the manipulation of heat conduction are based on microscopic mechanisms. At macroscale, there are relatively less accessible degrees of freedom, so designing the spatial distribution of thermal conductivity (including the geometric boundaries and interfaces) constitutes the majority of present passive approaches. Conductive thermal metamaterials[2,3] (Box 1) are artificial structures of such kind. They are often simply called thermal metamaterials, but here this term refers to all kinds of metamaterials for heat transfer control, including the nanophononic and photonic metamaterials that will be discussed in the following.

___

**Box 1 | Macroscopic heat conduction and conductive thermal metamaterials**
*Heat conduction equation*
Heat flows spontaneously from a high temperature region toward a low temperature region. According to the Fourier's law, the macroscopic heat conduction follows $\boldsymbol{q} = -\boldsymbol{\kappa}\nabla T$ where $\boldsymbol{q}$ is the heat flux density, $T$ is temperature, and $\boldsymbol{\kappa}$ is the thermal conductivity tensor. Combining with energy conservation, the temperature field follows

$$\rho c_p \partial T / \partial t = \nabla \cdot (\boldsymbol{\kappa} \cdot \nabla T) + g(\boldsymbol{r},t) \qquad (1)$$

where $t$ is time, $\boldsymbol{r}$ is the position vector, $g(\boldsymbol{r},t)$ is the rate of energy generation, $\rho$ and $c_p$ represent mass density and specific heat capacity at constant pressure, respectively. By ignoring interfacial thermal resistance[23], at the interface of two regions A and B with thermal conductivity $\kappa_A$ and $\kappa_B$, the temperature ($T_A$, $T_B$) and the heat flux normal to the interface should match

$$\begin{aligned} T_A &= T_B \\ -(\boldsymbol{\kappa}_A \cdot \nabla T_A) \cdot \boldsymbol{n} &= -(\boldsymbol{\kappa}_B \cdot \nabla T_B) \cdot \boldsymbol{n} \end{aligned} \qquad (2)$$

where $\boldsymbol{n}$ is the unit vector normal to the interface. Note that at nanoscale, the interfacial thermal resistance is nonnegligible and even asymmetric, leading to the asymmetric temperature drop at the interface and thermal rectification[24]. The effect of interfacial thermal resistance demands more attention as thermal devices come to nanoscale. Due to a limited length, we have to avoid discussion on this very broad topic which is more weighted on the microscopic mechanisms. Interested readers may refer to related review articles[23,25,26,27].

*Thermal conductivity*
Conventional materials often have uniform and isotropic thermal conductivities in the range of ~0.03 (air, expanded polystyrene) to ~400 W/m·K (copper, silver). The thermal conductivity of diamond[28] can reach ~2000 W/m·K, comparable to the in-plane thermal conductivity of bulk graphite[29], but they have not been commonly used due to the high cost of diamond and the anisotropy of graphite. Recently, there are some reports on the high thermal conductivities (~1500 W/m·K) of semiconductors like boron arsenide[30,31,32] and isotope-enriched boron nitride[33]. The thermal conductivity of most natural materials is isotropic ($\kappa$ is a scalar), but for single-crystalline materials such as graphite, the intrinsic orientation of the lattice could induce anisotropy. When the thermal conductivity is anisotropic, $\boldsymbol{\kappa}$ should be symmetric for common materials according to the Onsager reciprocity theorem[34]. When magnetic fields are present, the effective thermal conductivity of a material could be asymmetric because of the thermal Hall effect[35].

*Conductive thermal metamaterials*
As an information carrier, temperature field is measured in various thermography techniques[36]. It can also be operated for computations[4]. Conductive thermal metamaterials can manipulate temperature field with good robustness and accuracy. They thus provide interesting possibilities including the encryption[37], concealment[8,9], camouflage[38,39], and rectification[40] of heat signals measured by an infrared camera or a temperature probe. These effects have many direct and indirect applications. The camouflage of radiative signals can be directly used for infrared anti-detection. The concealment and processing of conductive heat signals are important building blocks for some potential applications. For example, a thermal concentrator[41] can collect heat from the environment for energy conversion. An energy-free thermostat[42] was proposed based on the design of thermal cloak. A modular design of thermal unit cells[43] can feasibly manage the temperature field on a circuit board. They bring attractive features like low-cost, adaptiveness, and tunability into the thermal energy utilization and temperature management, and are expected to help in realistic situations when cooperated with thermoelectric modules, heat sinks, etc.

___

***Transformation theory.*** Maxwell's equations elucidate how light propagates, if we know the permittivity and permeability distributions in space. Transformation optics enables us to solve the inverse problem, that is, how to realize a specific light field by designing the material parameters[45,46]. It is based on the form invariance of Maxwell's equations under coordinate transformation, which preserves identical solutions in two coordinates. One can thus realize an electromagnetic field by transforming from a known solution, such as a plane wave in free space. The material parameters are then derived from those of the free space in the new coordinate system. The transformation theory has been successfully applied on macroscopic heat conduction, thanks to the form invariance of the governing equation at steady states[8,9] and transient states[47]. For a transformation from a virtual space $r'$ to the real space $r$, the parameters in the two spaces are connected through the relationship[3] $\kappa = A\kappa'A^T/\det(A)$ and $\rho c_p = \rho'c_p'/\det(A)$, where $A$ is the Jacobian matrix for transformation $r(r')$.

One example is the thermal cloak that prevents heat from entering the cloaking region to keep a uniform interior temperature without perturbing the exterior field. The corresponding transformation in polar coordinate system $(r,\theta)$ is carried out along the radial direction with[47] $r = R_1 + (R_2 - R_1)r'/R_2$, where $R_1$ and $R_2$ are the interior and exterior radii of the cloak, respectively. Under it, region ($0 \leq r' \leq R_2$) is squeezed into a shell ($R_1 \leq r \leq R_2$), while the rest region ($r \geq R_2$) is unchanged. Heat flux is thus prevented from entering the cloaking region ($r \leq R_1$) while the exterior fields are not distorted. The material parameters of a cylindrical cloak has components: $\kappa_{rr}/\kappa_0 = \kappa_0/\kappa_{\theta\theta} = (r - R_1)/r$, $\kappa_{r\theta} = \kappa_{\theta r} = 0$, and $\rho c_p = \rho_0 c_{p0}[R_2/(R_2 - R_1)]^2(r - R_1)/r$, where $\kappa_0$, $\rho_0$, and $c_{p0}$ are the thermal conductivity, density, and specific heat capacity of the background, respectively.

These parameters have singularities that are difficult to be realized, so a reduced version was proposed[47]: $\kappa_{rr} = \kappa_0[R_2(r - R_1)/r(R_2 - R_1)]^2$, $\kappa_{\theta\theta} = \kappa_0 R_2^2/(R_2 - R_1)^2$, $\kappa_{r\theta} = \kappa_{\theta r} = 0$, and $\rho c_p = \rho_0 c_{p0}$. The reduced thermal cloak has no singularity in $\kappa_{rr}$, with constant $\kappa_{\theta\theta}$ and $\rho c_p$. Using them, the first transformation-based thermal cloak was experimentally demonstrated[48] (Fig. 1a). For practical realization, one needs to discretize the cloaking shell into $N$ homogenous layers. Then each layer, which is homogeneous but anisotropic, can be realized by drilling holes into a copper plate and filling them with PDMS (Fig. 1a) according to the effective medium theory[49]. An earlier work experimentally demonstrated a multilayered thermal cloak with only two kinds of natural bulk materials (homogeneous and isotropic)[10]. To realize an ideal thermal cloak based on transformation remains challenging. Existing devices usually have nice performance at steady state, but detectable in the transient state[50] due to the reduced density and heat capacity.

In contrast to thermal cloak, a thermal concentrator enhances its interior temperature gradient[10,47]. The corresponding transformation maps regions $R_3 \leq r' \leq R_2$ and $r' \leq R_3$ in the virtual space to regions $R_1 \leq r \leq R_2$ and $r \leq R_1$ (where $R_1 < R_3$), respectively. It can be realized by using two different materials aligned in the radial direction[41]. Another kind of heat collector is a thermal converging device that guides heat toward a desired spot by performing regional rotation transformations[51]. For both thermal cloak and concentrator, the region $r \geq R_2$ is unchanged, so one can easily put multiple copies of them in one temperature field without any interference. It was proposed to treat the zero and enhanced temperature gradients inside these independent devices as binary digits, such that information can be encoded by a temperature field[52]. Thermal cloaking and concentrating can also be realized within a doublet thermal meta-device[53], which consists of mechanically movable blocks of two materials.

Transformation theory is not only applicable to normal objects, but also to heat sources. Thermal illusion[38] is an effect where the temperature field surrounding a heat source is modified such that one who measures the new field will not perceive the real shape, size, or the number of the heat sources. It is achieved by transforming the region occupied by the heat source into other shapes.

According to the Stefan-Boltzmann law (see Section III), the surface temperature field and emissivity of an object determine its thermal emission and thereby its thermal image. Therefore, one can modify the radiative signal by manipulating the heat conduction near the surface of an object. Based on the idea, a structured thermal surface has been made for radiative camouflage[39,54]. Using a non-invasive transformation, the device can restore arbitrary background temperature distributions on its top. When the device has similar integrated emissivity as the background (which can be easily achieved by covering a thin film), its image measured by a common infrared camera will be almost identical to that of a pure background. Thus, it is expected to manipulate thermal radiation in complex environments. A big advantage of the transformation theory is that the function of the device is independent of the temperatures of the object and background, while emissivity engineering usually requires such knowledge. However, the devices often have bulky and rigid metal structures, which may not be conveniently applicable and require further optimization. The idea of using thermal conduction to manipulate the surface temperature and thereby the radiation was also demonstrated to generate arbitrary infrared thermogram without relying on transformation[55].

***Directly solving the equation.*** An object can be identified based on its unique scattering signature in various physical fields. Based on the scattering solutions, a DC magnetic cloak that consists of two layers of ferromagnetic and superconductor has been illustrated[56]. This has inspired the experimental realization of bilayer thermal cloaks in three-dimension[57] and two-dimension[58] (Fig. 1b). A bilayer thermal cloak consists of an inner layer ($R_1 < r < R_2$) and an outer layer ($R_2 < r < R_3$) with conductivities of $\kappa_2$ and $\kappa_3$, respectively. The inner layer is assumed to be a perfect thermal insulator ($\kappa_2 \to 0$), which ensures that external heat flux do not penetrate inside it. Then the outer layer must eliminate the external-field distortion. Directly solving the thermal conduction equation gives $\kappa_3 = \kappa_0(R_3^3 + 0.5R_2^3)/(R_3^3 - R_2^3)$ for spherical cloak and $\kappa_3 = \kappa_0(R_3^2 + R_2^2)/(R_3^2 - R_2^2)$ for cylindrical cloak, respectively. $\kappa_0$ is the thermal conductivity of the background. The third parameter can be uniquely determined if any two of $\kappa_3$, $\kappa_0$, $R_3/R_2$ are known. A different bilayer cylindrical device was recently proposed[22]. It has an extremely conductive inner layer $\kappa_2 \to \infty$, such that heat is quickly transported through it, irrespective of the interior object. This mechanism is reminiscent of the tunneling effect of near zero-index materials in photonics, and a rigorous correspondence has been proved. The outer layer is also required to restore distortion with $\kappa_3 = \kappa_0(R_3^2 - R_2^2)/(R_3^2 + R_2^2)$. Thanks to the unique mechanism, the interior object has good sensitivity to environmental changes.

Inspired by the bilayer scheme, multiphysics cloak[59] and invisible sensor[60], which can manipulate heat flux and electric current simultaneously, have been experimentally realized. By modifying the bilayer structure, thermal camouflaging has been demonstrated to create multiple images off the original object's position in heat conduction[61]. Based on the scattering theory, thermal transparency[62] and passive metashells with adaptive thermal conductivities[63] have also been proposed. By directly solving the conduction equation in the elliptical coordinate system ($\xi$, $\eta$), the bilayer strategy has been successfully applied to the design of an elliptical cloak, with an assumption of anisotropic background[64]. Instead of analytically solving the governing equations, optimization algorithms can be used to obtain the structure that gives certain temperature distribution, as demonstrated in a topology-optimized thermal carpet cloak[65]. When different materials are combined to realize bilayer or multi-layer thermal devices, interfacial thermal resistance is a practical issue that may influence their performances[66].

***Temperature dependence.*** The thermal conductivities of most materials are temperature dependent. For some phase-change materials this temperature dependence is drastic and can be used to build thermal diodes, regulators, and switchs[67]. Traditional transformation theories[8,9,47] only applies to linear materials whose thermal conductivity is independent of temperature. This issue was resolved by proving that the temperature-dependent Fourier's law still keeps form invariance under coordinate transforamtion[40,68]. Therefore, the nonlinear thermal conductivity in the real space is obtained as $\kappa(T) = A\kappa'(T)A^T/\det(A)$, which holds the same form as for temperature-independent cases[8,9,47]. Furthermore, it is proposed that the equation can be equivalently written as $\kappa(T) = A(T)\kappa_0 A(T)^T/\det[A(T)]$, where $A(T)$ is a temperature-dependent transformation. This method can be utilized to design devices such as switchable thermal cloak[40] and concentrator[69]. A macroscopic thermal diode was also designed and fabricated using parts of the switchable thermal cloak[40] (Fig. 1c). In addition, temperature-dependent transformation is also used to design multifunctional devices, such as thermal cloak-concentrator[70]. Without relying on transformation, an energy-free thermostat was designed by studying the temperature-dependent one-dimensional thermal conduction[42], where a temperature trapping effect was found that maintains near constant temperature for various boundary conditions. It can be used for the temperature managements of targets in large temperature differences.

Strongly temperature-dependent thermal conductivity cannot easily be found in natural materials. Therefore, the experimental demonstration of above devices all rely on the large mechanical deformation of shape-memory alloys (SMA) across phase transition, which effectively modifies the thermal conductivity by changing the contact area between components. SMA is widely used to make thermal swithes and thermal diodes[67], thanks to its moderate and tunable phase transition point as well as configurable deformation patterns. One interesting application is the thermal management using SMA-based thermal regulator that increases the safty and capacity of Li-ion batteries[71]. Phase change nanocomposite is also as an effective tool. Based on it, a modular design of temperature-responsive thermal metamaterial has been demonstrated to manage the temperature field on a circuit board with switchable heat shelding effects[43].

**Between macro- and microscale.** Despite of sharing the same term, conductive thermal metamaterials are different from other kinds of metamaterials in many aspects. First, they manipulate diffusion instead of wave propagation. Second, there is no sub-wavelength structure in them. Finally, since there is no such mechanism as local resonance and their effective material parameters (thermal conductivity, mass density, heat capacity) are not easily adjustable, letting alone the realization of unconventional values (negative or zero). It seems that some common features of metamaterials are lacking, which simplifies the design and fabrication but restricts the performance and applications. A major challenge is to break the restrictions and discover new mechanisms for the control of macroscopic heat conduction.

Some opportunities may exist at microscale. For example, a recent work shows that the thermal conductivity of silicon nanowire can be locally tuned to any desired value between 2 and 50 W m$^{-1}$ K$^{-1}$ with a suitable dose of helium ion irradiation[72], which introduces point defects and triggers crystalline-amorphous transition beyond a threshold. Moreover, the defects can be annealed at higher temperatures, making the engineering process reversible to some extent. A continuously addressable thermal conductivity profile can thus be built on a single material while maintaining a uniform morphology. It is more feasible than the discrete and fixed distributions in most conductive thermal metamaterials. As a demonstration, bilayer and trilayer thermal cloaks have been fabricated on a 20 × 20 μm silicon membrane based on the ion-write mechanism[73] (The Fourier's law is still valid at the length scale).

More microscopic mechanisms for thermal conductivity engineering will be discussed in the following. Also, many active sources such as electric fields, pressure, and chemical stimulation enable reversible and even dynamic tuning of thermal conductivity (see Section IV). However, among them little effort has been made to engineer a thermal conductivity with spatial variations. Future developments in this direction may benefit from existing designs of conductive thermal metamaterials, but still face the challenges of mass production and accurate driving sources.

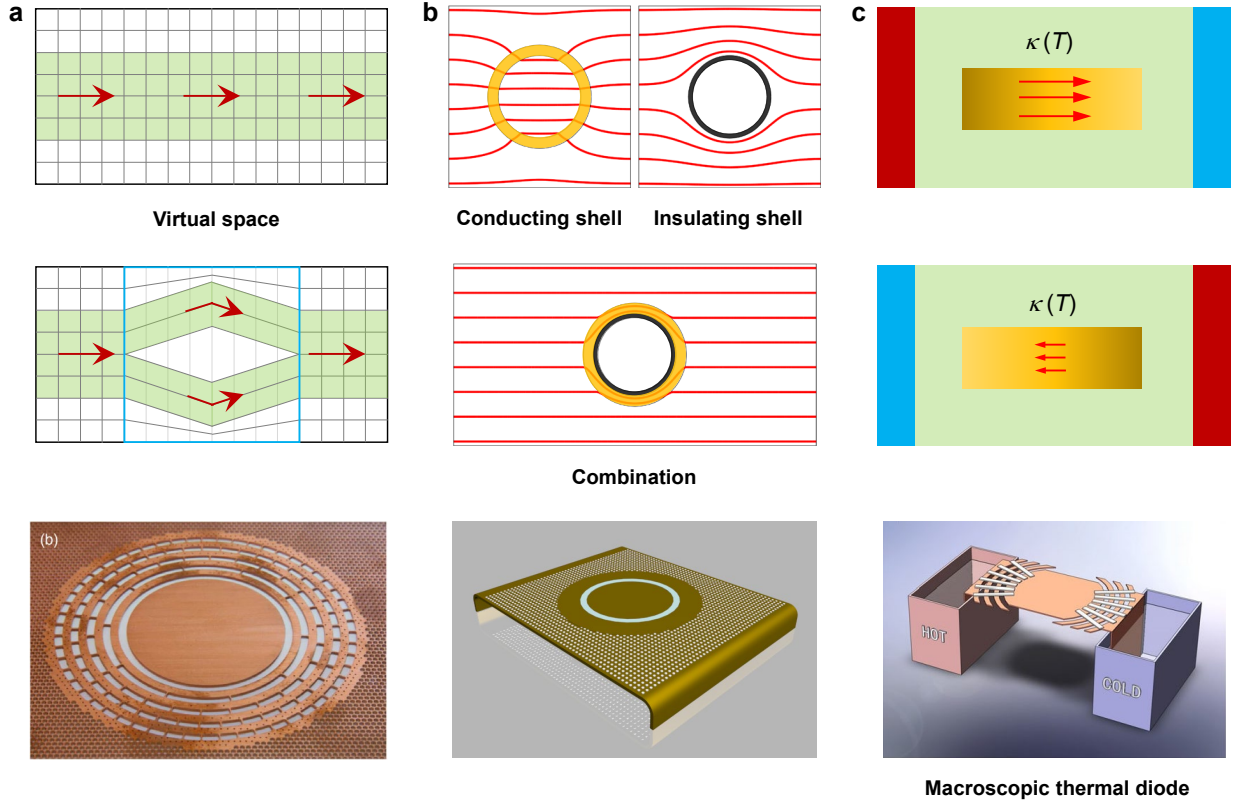

**Figure 1 | Macroscopic approaches for manipulating heat conduction. a |** Transformation theory makes the real space equivalent to the virtual space when a thermal metamaterial is filled in the transformation region, which provides a powerful tool for the design of transformation-based thermal cloak[48]. **b |** Direct solving the equation results in a bilayer cloak that consists of an inner insulating shell and an outer conducting shell[61]. **c |** Temperature-dependent transformation creates the devices with asymmetric behavior of the heat current, e.g. thermal diode conducts heat in one direction but insulates the heat in the opposite direction[40].

**Microscopic approaches**

Phonon, a kind of quasiparticle representing the collective vibrational mode of materials, is one of the main heat carriers especially in semiconductors and dielectric materials. An emerging field termed nanophononics aims to manipulate thermal properties by controlling phonon behaviors. Based on the phonon Boltzmann transport equation (BTE), thermal conductivity under single mode relaxation time approximation can be expressed as

$$\kappa = \frac{1}{24\pi^3}\sum_n \int C(\boldsymbol{k},n,T) v_g^2(\boldsymbol{k},n) \tau(\boldsymbol{k},n,T) d\boldsymbol{k} \qquad (3)$$

where $n$ is dispersion branch, $\boldsymbol{k}$ is the phonon wave vector, $T$ is temperature, $C$ is phonon specific heat of each mode, $v_g$ is group velocity and $\tau$ is phonon lifetime. The group velocity $v_g$ is defined as the derivative of phonon dispersion: $v_g = \partial\omega/\partial k$. The summation over dispersion branch $n$ can also be replaced by an integral over frequency $\omega$ with an additional factor termed density of state (DOS).

A widely used approach to manipulate the heat conduction is to modify phonon lifetime, which can be realized by doping, changing surface roughness, etc. Another potential approach is to modify phonon properties by the nanostructured devices, which is challenging due to the short wavelength of dominant thermal phonon and has attracted intense interests amid the recent great progress in nanofabrication. The wave nature of phonon sparked two active subjects: nanophononic crystals and nanophononic metamaterials.

***Nanophononic crystals.*** Phononic crystals (PnCs) are artificial periodic nanostructures whose periodicity is on the order of the phonon wavelength[74]. To distinguish PnCs that manipulate thermal phonons from those that influence elastic waves, the former are usually referred to as nanophononic crystals (NPC). NPC can be one-dimensional (1D, superlattices, Fig. 2a), two-dimensional (2D, nanomesh, Fig. 2b), or three-dimensional (3D, Fig. 2c). The wave interference induced by the periodic structures can give rise to many novel phenomena that cannot be found in conventional bulk materials, such as Brillouin zone folding, bandgap opening, etc. There are three important length scales in NPC with smooth boundaries (Fig. 2b): phonon wavelength ($\lambda$), phonon mean free path (MFP, $\Lambda$) and periodicity of the phononic crystal ($d$). According to the size relationship, the phonon transport can be categorized into three regions: (i) when periodicity is smaller than phonon wavelength ($d < \lambda$), wave-like phonon interference induced by coherent scattering (Bragg diffraction) is dominant; (ii) when periodicity is larger than phonon MFP ($d > \Lambda$), particle-like diffusive scattering is dominant; (iii) when periodicity is larger than phonon wavelength but smaller than phonon MFP ($\lambda < d < \Lambda$), however, the relative contribution of phonon interference and diffusive scattering to the heat transport is inconclusive. It is also worth pointing out that in this region there is a portion of phonons undergoing particle-like ballistic transport, which has motivated the concept of ray-like heat manipulation in phononic nanostructures, such as directional phonon source and thermal lens[75,76,77].

For 1D NPC, it was observed that thermal conductivity in a superlattice of W/Al$_2$O$_3$ can be 4 times smaller than the series average of the conductivities of the alumina and W layers[78]. The crossover from diffusive to coherent phonon scattering, marked by the minimum of thermal conductivity as a function of period thickness, was first predicted in theory[79,80] and then observed in epitaxial oxide superlattices[81] (Fig. 2a). For 3D NPC, extreme low thermal conductivity was predicted in silicon crystal with spherical pores, even at temperature up to 1100 K[82,83,84] (Fig. 2c). For 2D NPC, a large reduction in thermal conductivity was observed in a wide temperature range, from 0.1 K to 300 K[13,85,86,87,88,89,90,91,92,93,94]. The remarkable reduction in 3D and 2D NPC cannot be simply explained by the porosity correction factor that describes the bulk porous materials[83,85]. Next we will focus on the case of 2D NPC because it is actively investigated in experiments.

At low temperature, the effect of phonon interference is manifested through the strong difference in the temperature dependence of thermal conductance between the full membrane and the two 2D NPC samples[88] (Fig. 2b). At room temperature, the role of phonon interference is still in debate. Both phonon interference and diffusive scattering can decrease the thermal conductivity. For phonon interference, the reduction originates in the decrease of group velocity $v_g$ as a consequence of the flattening of the phonon band structures[13,86,87,88,95]. For diffusive scattering, the reduction is due to the decrease of phonon MFP and lifetime[85,86,87,92,93,95]. In experiments, methods of distinguishing these two mechanisms include studying thermal conductivity of NPC with fixed periodicity but different unit cell geometries[89,91], periodic and aperiodic nanomesh with the same porosity[93,94], etc. But no unambiguous conclusion is reached. In theory, both partially coherent scattering model[89,95] and diffusive scattering model[93,96,97] can show good agreement with experiments.

The inconsistence comes from the insufficient knowledge of the coherent length scale. Detecting phonon interference by measuring thermal conductivity is an indirect methodology. Due to the short wavelength of thermal phonon, a small difference in the geometry or surface roughness of NPC can greatly modify the phonon coherence, which may lead to the ambiguous conclusions in different experiments. A direct detection of phonon coherence can provide more information. Recently, the use of two-photon interference such as coherent population trapping (CPT) or electromagnetic induced transparency (EIT) to detect phonon coherence was proposed[98].

***Nanophononic metamaterials.*** Nanophononic metamaterials (NPM) use the idea of local resonances to manipulate heat conduction. The reduction of thermal conductivity in NPM was first revealed in pillar based structures[14]. A further study of the optimal size and geometry of NPM showed a two orders of magnitude reduction[99,100] (Fig. 2d). Recently, the concept of NPM has also been applied to nanoribbons[101] and nanotubes[102]. External periodic pillars acting as local resonators are deposited on base materials, introducing additional flat branches with resonance frequencies to the phonon dispersion of the base materials. The hybridization between the local resonances of pillars and the vibrational modes of base materials induces avoided crossing in the phonon dispersion, opening bandgaps and flattening the original branches near the resonance frequencies and hence lowering the group velocity and thermal conductivity. NPM have some potential advantages over NPC in the design of thermoelectric devices. The externally deposited pillars have a small effect on the electron transport in the base materials and will not decrease the device strength. What's more, periodic pillars can be treated as a NPC with a new degree of freedom. For the detailed analysis of geometry and size effect for NPM and a direct comparsion between NPM and NPC, we refer the interested reader to the recent comprehensive review[103].

Although the reduction of thermal conductivity has been observed in both experiments[104] and numeircal simulations[14,99,105,106,107,108,109], the role of local resonances is still in debate. The following three mechanisms are conjectured as possible mechanisms of the reduced thermal conductivity: (i) local resonance; (ii) phonon interference induced by coherent scattering and (iii) diffusive boundary scattering.

To remove the effect of phonon interference, researchers numerically studied the thermal conductiviy of aperiodically deposited pillar[106]. The large reduction of thermal conductivity indicates the small contribution of phonon interference but is unable to distinguish between the contribution of local resonance and diffusive boundary scattering. To isolate local resonance from diffusive boundary scattering requires fabricating pillars with no local resonances in the frequency region of dominant phonons, which has been rarely studied. Another approach to uncover the role of the three mechanisms is to study the dependence of thermal conductivity on the pillar height[99,105,108,109]. Molecular dynamics (MD) simulations showed that periodic pillars as short as a few angstroms can cause a drop of more than 50% in thermal conductivity[105,109]. In such short pillars, almost no local resonance frequencies are inside the frequency region of dominant phonons. This fact suggests that the existence of periodicity (phonon interference) contributes to the reduced thermal conductivity. As the pillar height increases, the results of MD simulations and finite element method (FEM) are inconsistent. While MD simulations showed that the reduction gradually saturates[105,109], FEM showed that the thermal conductivity will increase after a critial pillar height[108]. This inconsistency, which may be due to the lack of anharmonic phonon-phonon interaction in FEM, indicates the possible role of diffusive scattering.

The physical mechanism behind the reduced thermal conductivity of pillar-based NPM remains an open question. It's hard to isolate local resonance, phonon interference, and diffusive scattering in current experiments. As what we have illustrated in the discussion of NPC, a direct detection of phonon coherence can provide more useful information.

Last but not least, we would like to point out another kind of resonance due to the coupling of longituinal modes and transverse mode, which was used to reduce the thermal conductivity in core-shell nanowire[110].

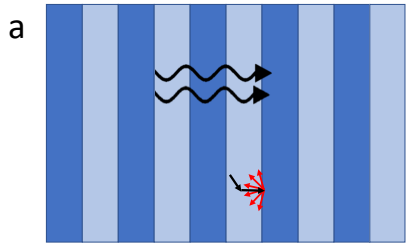
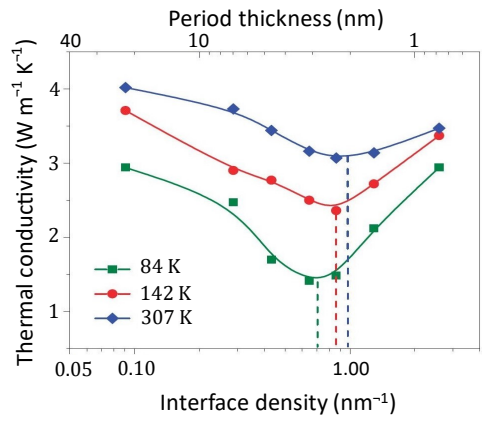

a

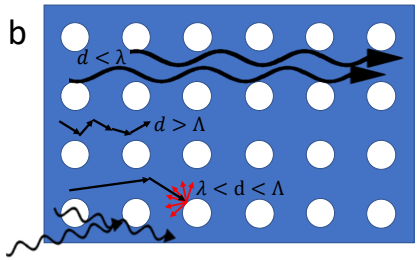
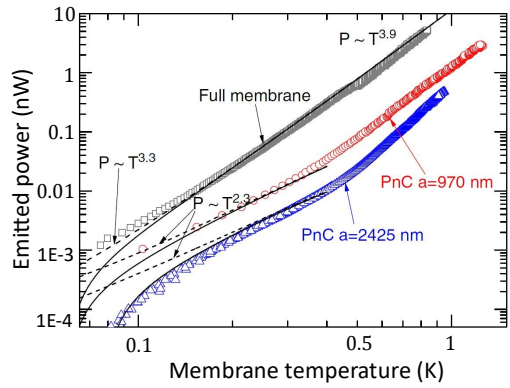

b

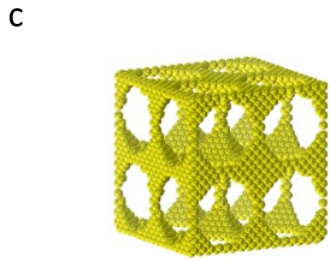
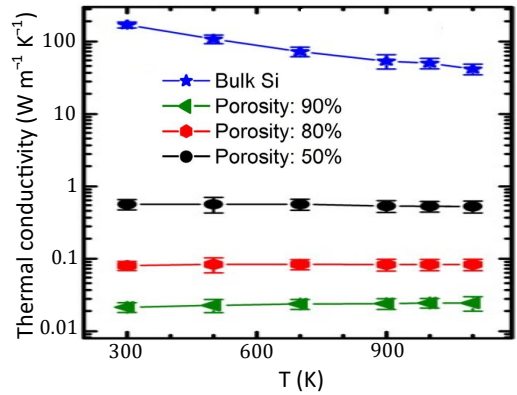

c

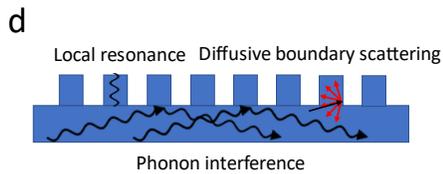
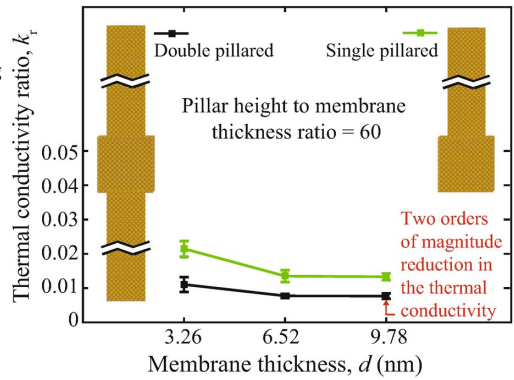

d

**Figure 2 | Microscopic approaches in heat conduction**. **a |** Left: coherent and diffusive phonon scattering in 1D superlattices. Right: Measured thermal conductivity of $(SrTiO_3)_m/(CaTiO_3)_n$ superlattices versus interface density at different temperatures. The minimum thermal conductivity indicates the crossover from diffusive to coherent phonon scattering. The minimum of thermal conductivity becomes deeper at lower temperature and the corresponding interface density moves to smaller values at lower temperature, which further corroborates the observation of coherent wave effect[81]. **b |** Left: Three regions of phonon transport decided by the relationship between phonon wavelength $\lambda$, phonon MFP $\Lambda$ and periodicity $d$ in 2D NPC. Right: Measured emitted phonon power versus temperature. Grey squares show the data for the full membrane, red circles (periodicity a = 970 nm) and blue triangles (periodicity a = 2425 nm) show the data for the two square 2D NPC samples. Theoretical calculations with and without a phonon back radiation power from the substrate are shown by solid and dashed lines, respectively. At low temperature, the temperature dependence of conductance is different in the full membrane ($P \sim T^{3.3}$) and NPC ($P \sim T^{2.3}$) cases[88]. **c |** Left: 3D NPC. Right: Thermal conductivity of 3D NPC and bulk Si versus the temperature. When the porosity is 90%, the thermal conductivity predicted from the classical Eucken model is about 12 W m$^{-1}$ K$^{-1}$, which is two orders larger than that of 3D NPC with porosity of 90%[83]. **d |** Left: Three mechanisms that induce the reduced thermal conductivity in pillar-based structure. Right: Thermal conductivity reduction of the double pillared membrane and the single pillared membrane compared to a uniform membrane with the same thickness[100]. A reduction by two orders of magnitude is achieved.

## III. Manipulating thermal radiation

Thermal radiation represents a ubiquitous aspect of nature and a fundamental heat transfer mechanism[111,112,113]. Any object at finite temperatures emits thermal radiation due to the thermally induced motion of particles and quasiparticles. Conventional thermal emitters have a set of common characteristics[113] (Fig. 3a) and fundamental constraints (Box 2), which impose strong restrictions on their capabilities for controlling heat transfer processes.

Recent advances in using sub-wavelength structures have led to properties that are drastically different from those of the conventional structures. Moreover, when bringing objects to the near field regime, in which the gap distance is comparable or smaller than the thermal wavelength, the radiative heat transfer can be significantly larger than that in the far field. These approaches provide new capabilities for controlling the fundamental aspects of thermal radiation, and open up new opportunities for applications. Here we give a brief overview of the fundamental advances in manipulating both far field and near field thermal radiation, as well as a few highlights of applications.

___________________________________________________________________________________

**Box 2. Fundamental properties of thermal radiation**
**a | The conventional view on fundamental properties and constraints of thermal radiation**
*Planck's Law*
The thermal emission power $P_0$ of a blackbody emitter in thermal equilibrium at a temperature $T$, emitting into free space, can be described by Planck's Law:

$$P_0 = A \cdot \frac{\omega^2}{4\pi^2 c^2} \cdot \frac{\hbar\omega}{e^{\hbar\omega/k_B T} - 1} \tag{4}$$

Here $A$ is the emitter area, $\omega$ is the angular frequency of the emission, $\hbar$ and $k_B$ are the reduced Planck constant and the Boltzmann constant, respectively. $c$ is the speed of light in vacuum.

*Absorptivity and emissivity*
The key quantities characterizing thermal emitters in the far field are the angular spectral absorptivity $\alpha(\omega, \hat{n}, \hat{p})$, and the angular spectral emissivity $e(\omega, \hat{n}, \hat{p})$. $\alpha(\omega, \hat{n}, \hat{p})$ represents the absorptivity of the emitter for incident light at a frequency $\omega$ and direction $\hat{n}$ with a polarization vector $\hat{p}$, and is measured as the ratio between the incident and the absorbed power per unit area. $e(\omega, \hat{n}, \hat{p})$ measures the spectral emission power per unit area at certain $\omega$, $\hat{n}$, and $\hat{p}$, normalized against the spectral emission power per unit area of a blackbody emitter at the same $\omega$, $\hat{n}$, and $\hat{p}$. Thus for blackbody, both $\alpha(\omega,\hat{n},\hat{p})$ and $e(\omega,\hat{n},\hat{p})$ are equal to 1. For typical macroscopic thermal emitters $\alpha(\omega,\hat{n},\hat{p})$ and $e(\omega,\hat{n},\hat{p})$ are between 0 and 1.

*Kirchoff's Law*
The angular spectral absorptivity and emissivity typically satisfy Kirchoff's Law:
$$\alpha(\omega,\hat{n},\hat{p}) = e(\omega,\hat{n},\hat{p}^*) \tag{5}$$
Here the $\hat{p}^*$ denotes complex conjugation. Equation (5) is valid for reciprocal system consisting of materials characterized by symmetric permittivity and permeability tensors.

**b | Far-field and near-field radiative heat transfer**
Taking the simple example of two semi-infinite parallel plates separated by a vacuum gap with a size $d$, the radiative heat flux density $q$, or the radiative heat flux across the gap per unit area is[114]
$$q = \int_0^\infty \frac{d\omega}{4\pi^2} \left[\Theta(\omega,T_1) - \Theta(\omega,T_2)\right] \int_0^\infty dk \left[\xi_s(\omega,k) + \xi_p(\omega,k)\right] k \tag{6}$$
where $\Theta(\omega,T) = \hbar\omega/[\exp(\hbar\omega/k_B T) - 1]$, $k$ is the wave vector parallel to the planar surfaces. $\xi_s(\omega,k)$ and $\xi_p(\omega,k)$ are the photon transmission probability for the transverse electric ($s$) and transverse magnetic ($p$) polarizations, respectively, and can be written as:
$$\xi_{s,p}(\omega, k < k_0) = \frac{\left(1-|r_1|^2\right)\left(1-|r_2|^2\right)}{\left|1-r_1 r_2 e^{-2ik_{z0}d}\right|^2}, \text{ for propagating waves} \tag{7}$$

$$\xi_{s,p}(\omega, k > k_0) = \frac{4 Im(r_1) Im(r_2) e^{-2|k_{z0}|d}}{\left|1-r_1 r_2 e^{-2ik_{z0}d}\right|^2}, \text{ for evanescent waves} \tag{8}$$

In equations (7) and (8), $k_0 = \omega/c$ is the free space wavenumber, $k_{z0}$ is the out of plane wavenumber in vacuum. The parameters $r_{1,2}$ are the Fresnel reflection coefficients for incidence from the gap to bodies 1 and 2. Both $\xi_s$ and $\xi_p$ are between 0 and 1. In most radiative heat transfer cases with non-magnetic materials, $\xi_p(\omega,k)$ dominates. As a result, the summed photon transmission probability $\xi(\omega,k) = \xi_s(\omega,k) + \xi_p(\omega,k)$ is typically between 0 and 1.

In the far field regime, only the propagative waves (equation (7)), contribute to the total heat flux. In the near field regime, when the gap size is smaller than the characteristic thermal wavelength, evanescent waves (equation (8)) also contributes (Fig. 4a). As a result, the flux density in the near field regime can exceed that in the far-field regime by orders of magnitude.

__________________________________________________________________________________

**Far field thermal radiation control**
*Fundamental properties control.* In nanostructures where the feature sizes are comparable to thermal radiation wavelength (Fig. 3b), one can design the structure to control all aspects of the the emissivity $e(\omega,\hat{n},\hat{p})$ [15,115,116,117,118,119,120,121,122]. As examples to control the spectral characteristics, one can strongly suppress thermal emissivity at a certain frequency using photonic band gap[115,116]. One can also strongly enhance the emissivity of a material, with a variety of photonic resonators[117,120]. A lossy resonator can create a spectrally sharp absorption peak and therefore a narrowband thermal emission peak. Moreover, by constructing multiple resonances, strong thermal emission with multiband or broadband characteristics can be achieved[118,120,122]. As examples to control angle and polarization characteristics, strong angular dependent thermal radiation has been demonstrated from a SiC grating[15]. One can also achieve strongly polarized thermal radiation, by constructing polarization-dependent resonances. Thermal emission with linear, circular and arbitrary polarization can be constructed, with anisotropic, chiral, and spin-optical metasurfaces[119].

Planck's law (equation (4)) sets an important constraint in far field thermal radiation. A photonic structure can have drastically different absorption and geometric cross-sections. Super-Planckian radiation can be realized when the absorption cross-section exceeds the geometric cross-section, as has been demonstrated in microscopic thermal antennas[123,124,125] as well as in macro-scopic emitters using a thermal extraction scheme[126] (Fig. 3c). Efforts have also been made toward the extraction of near field thermal radiation into far field[126,127,128].

Non-equilibrium thermal emitter offers possibilities of radiation control beyond equilibrium systems. One example of a non-equilibrium thermal emitter is an electrically biased or optically excited semiconductor, within which the electrons and the holes can have different quasi-fermi levels and therefore a non-zero chemical potential $\mu_\gamma$ (Fig. 3d). Such a semiconductor emits with a spectral energy density $\rho(\omega)$[129]:

$$\rho(\omega) = \frac{\omega^2}{\pi^2 c^3} \cdot \frac{\hbar\omega}{e^{(\hbar\omega-\mu_\gamma)/k_B T} - 1} \tag{9}$$

The Planck distribution is recovered at $\mu_\gamma = 0$. At the same emitter temperature, a positive chemical potential leads to $\rho(\omega)$ significantly beyond Planck's distribution. On the other hand, a negative chemical potential leads to a significantly suppressed $\rho(\omega)$. The manipulation of photon chemical potential enables new opportunities in heat manipulation using optoelectronic technology, as we will discuss further in the near field thermal radiation section.

Non-equilibrium thermal emission also occurs from a temporally modulated thermal emitter[130], with modulation time scale significantly smaller than the thermal time constant. Using such an emitter, heat can be actively pumped from a cold emitter to a hot emitter, resulting in a photon-based active cooling mechanism. A closely related approach to realize non-equilibrium thermal emitter is to explore the nonlinear effect[131,132] (Fig. 3e).

The vast majority of thermal emitters satisfy Kirchhoff's law (equation (5)) since most materials are reciprocal. This imposes significant constraints in solar energy harvesting: an efficient absorber will also be an efficient emitter, and therefore must re-emit part of the energy back to the sun, representing an intrinsic loss mechanism. The theromdynamic limit of solar energy harvesting can only be reached by breaking Kirchhoff's law[133,134]. Near-complete violation of the Kirchoff's law can be achieved by combining resonant nanostructure design with magneto-optical material[135] (Fig. 3f). Nonlinear or time varying permittivity/permeability can be used to break the balance between absorptivity and emissivity.

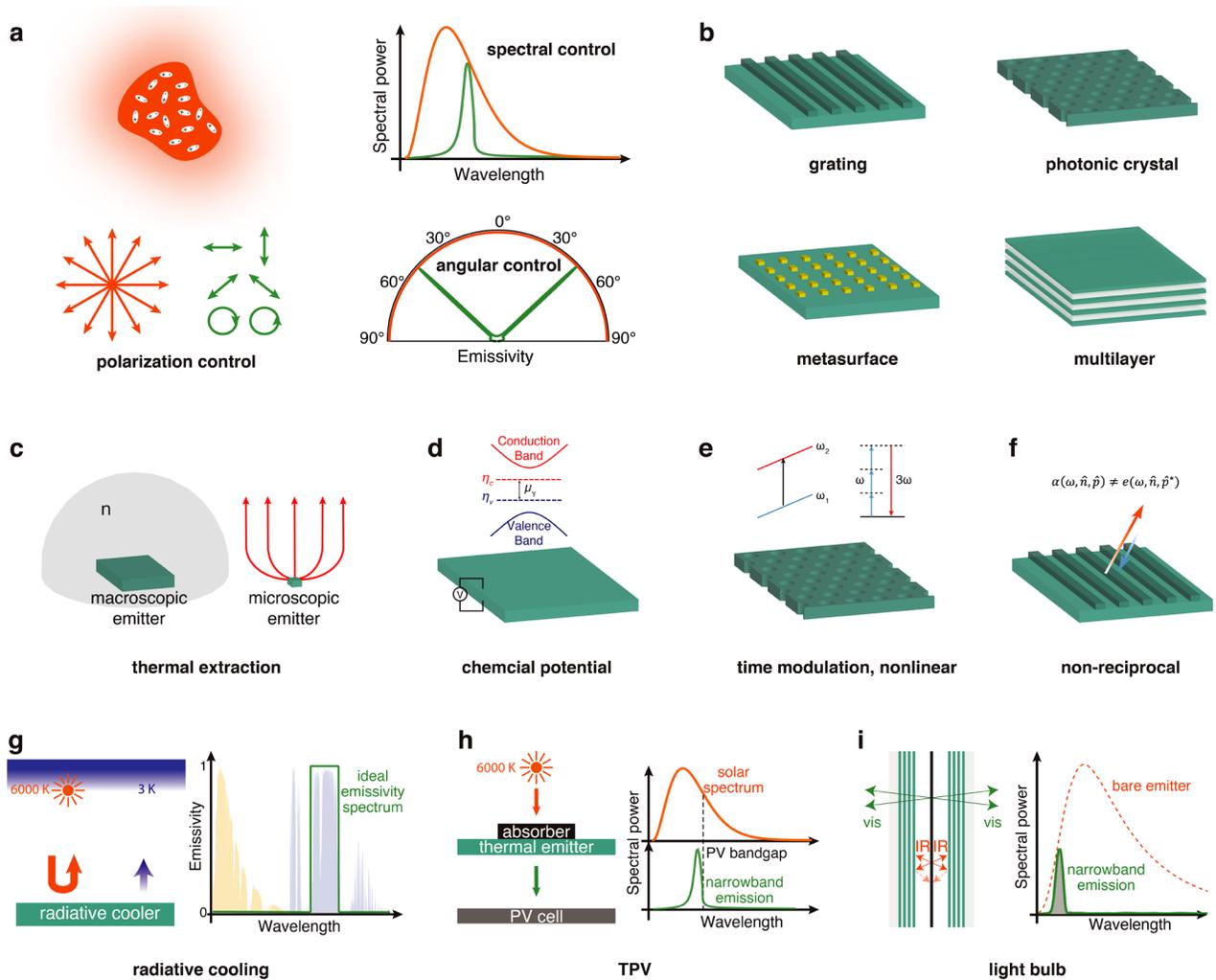

**Figure 3 | Far field thermal radiation manipulation. a |** Conventional blackbody thermal radiation (orange) and nanostructures controlled thermal radiation (green). **b |** Nanostructures for thermal radiation control, including photonic crystals, gratings, metamaterials, and multilayer films. **c |** Beyond Planck's Law: enhancing the absorption cross-section with thermal extraction on macroscopic emitters (left) and microscopic thermal antenna (right). **d |** Non-equilibrium thermal radiation with chemical potential $\mu_\gamma$, determined by the separation of quasi-Fermi levels for electrons $\eta_c$ and holes $\eta_v$. **e |** Non-equilibrium thermal radiation with time-modulation induced frequency transition and nonlinearity (for example, from $\omega_1$ to $\omega_2$ under time modulation, and from $\omega$ to $3\omega$ under third-order nonlinearity). **f |** Beyond Kirchhoff's Law: non-reciprocal thermal radiation. The balance between angular spectral absorptivity $\alpha(\omega, \hat{n}, \hat{p})$ and emissivity $\varepsilon(\omega, \hat{n}, \hat{p}^*)$ can be broken in non-reciprocal systems. **g |** Radiative cooling: daytime radiative cooling can be achieved with a near-zero absorptivity over the entire solar spectrum and a strong emissivity in 8–13 μm. **h |** Solar thermophotovoltaic system: an intermediate element absorbs incoming sunlight to heat up, and then generates thermal emission tailored to the solar cell bandgap. **i |** Incandescent light bulb: the efficiency can be significantly improved by suppressing the near-infrared emission while allowing the visible light emission.

*Applications.* The fundamental advances in far-field thermal radiation manipulation have led to many important application[5]. In this section, we highlight a few examples, including daytime radiative cooling, thermophotovoltaic system, and incandescent lighting.

The universe, at 3 K, represents the ultimate heat sink. Thermal radiation enables access to the cold universe: the blackbody radiation at typical ambient temperature has a large spectral overlap with the earth atmosphere transparency window in the wavelength range of 8–13 μm. Thus, any object on earth, given sky access, can radiate heat out to the universe, resulting in radiative cooling. Although nighttime radiative cooling has been known for centuries, daytime radiative cooling is more desirable and also more challenging due to the fact that the sun heats up most materials during the day. Achieving daytime radiative cooling requires a near-zero absorptivity over the entire solar spectrum and a strong emissivity in 8–13 μm (Fig. 3g). Daytime radiative cooler with this spectral feature has been recently theoretically proposed[136] and experimentally demonstated[137]. Since then, there have been significant progresses towards large scale deployment of this technology[138,139,140,141]. With its ability to harness the coldness of the universe and its broad implications for energy technology[142], radiative cooling has now opened up a new frontier in renewable-energy research[143].

In the area of solar energy harvesting, single-junction solar cells are subject to an efficiency limit of 41%, known as the Shockley–Queisser limit. A significant energy loss arises from the mismatch between the broadband solar radiation and the intrinsic properties of the single-junction solar cells. Solar thermophotovoltaic (STPV) systems represent an important avenue towards overcoming the Shockley-Quesser limit[144]. In a STPV system, an intermediate element absorbs incoming sunlight to heat up, and then generates thermal emission tailored to the solar cell bandgap (Fig. 3h), resulting a theoretical efficiency limit of 85%[145]. Proof of principle demonstration of such a concept has been reported in[146,147]. Closely related to STPV, a thermophotovoltaic (TPV) system can utilize thermal radiation from a local heat source to generate electricity in a photovoltaic cell[148]. In TPV systems, instead of suppressing below-bandgap radiation, one can alternatively do below-bandgap photo recycling, using photovoltaic cell with minimum below-bandgap absorption and high quality back reflector[148]. Further improvement on the photonic, thermal and carrier management of emitters and cells, manufacturing and integration of high quality materials may ultimately push STPV and TPV toward practical implementations.

In the area of incandescent lighting systems, conventional incandescent light bulb has low efficiency since its thermal radiation is mostly in the infrared wavelength range, and only a small portion produces visible light that can be used for lighting purposes. One way to reduce the loss is to significantly suppress the near-infrared emission while allowing the visible light emission (Fig. 3i). An experimental demonstration of this concept[149], shows efficiency approaching that of commercial fluorescent or light-emitting diode bulbs. The performance can be further improved with spectral engineering and optimization[150,151].

**Near-field radiative heat transfer (NFRHT) control**
As noted above, heat transfer in the near field can exceed the far-field blackbody limit by orders of magnitude (Box 2)[152-156]. In this section, we briefly review the recent advances and application opportunities in near-field thermal radiation.

***Efforts towards enhancing NFRHT.*** Based on equation (6), to enhance the overall heat transfer, one needs to maximize photon transmission probability $\xi(\omega,k)$ over all accessible frequency $\omega$ and parallel wave vector $k$ channels. The theoretical maximum heat transfer can be intuitively understood from an $\omega$–$k$ diagram (Fig. 4b). For two planar bodies with gap distance $d$, the maximum accessible parallel wave vector $k$ for photon transmission is about[157] $1/d$. This defines a region in the $\omega$–$k$ space within which the photon transmission channels are accessible (Fig. 4b). Such a treatment provides an intuitive understanding of the $1/d^2$ scaling of the magnitude of heat transfer coefficient in the near field regime, as well as an optimistic upper bound of the heat transfer coefficient[157].

To physically realize such large enhancement in NFRHT, one can choose materials that support surface waves such as surface phonon polaritons (SPhP) and surface plasmon polaritons (SPP) to access the large $k$ channels, for example, by using polar dielectrics such as SiC and $SiO_2$ that support SPhP[158], or doped semiconductors such as Si that support SPP[159]. A plot of photon transmission probability $\xi(\omega,k)$ between a pair of SiC with a gap distance of 10 nm is shown in Fig. 4c. With the surface wave excitation, a significant number of $\xi(\omega,k)$ channels can be accessed in the regime where $k$ can be orders of magnitude larger than $k_0$. On the other hand, we note that such a typical surface mode only uses a small fraction of the entire $\omega$–$k$ space, and hence limits the overall heat transfer enhancement.

To further enhance NFRHT, there have been significant efforts in exploring nanostructures and more complex geometries (Fig. 4e-i). The first approach is to carefully engineer the surface wave mode with correct frequency and loss. It has been shown[160] that the maximum heat flow can be achieved with spectrally matched emitters and receivers when the materials dielectric function $\varepsilon$ on both sides approach $-1 + i\alpha$, where $\alpha \ll 1$. Here, an important consideration is that ideally one should choose the surface wave mode excitation energy to be close to $k_B T$ where the thermal photon population is the largest. As a result, it has been shown[161] that Si-based metasurfaces (Fig. 4e) featuring two-dimensional periodic arrays of holes can exhibit a room-temperature NFRHT coefficient much larger than any unstructured material. Introducing holes in the Si layers reduces the losses and the effective plasma frequency that, in turn, redshifts the surface modes. This way, the surface modes are more easily occupied at room temperature, leading to NFRHT enhancement.

The second approach is to increase the number of resonant modes using multi-resonance nanostructures. Such an enhancement can be schematically visualized in Fig. 4d, where multiple modes can be constructed to fill up a larger regime in the $\omega$–$k$ space. Metasurfaces, photonic crystals[162], grooves[163], and multilayer[164] structures can all be designed for this purpose (Fig. 4g,h). For example, it has been shown that NFRHT between multilayer thermal bodies (Fig. 4h) can be significantly enhanced by surface states at multiple surfaces[164]. There are also efforts using hyperbolic dispersion for enhancing and broadening the filling in the $\omega$–$k$ space to enhance NFRHT[165].

Despite all the efforts so far, there are still two major open questions in understanding NFRHT. First, from the fundamental limit side, the NFRHT upper bound we have discussed (Fig. 4b, Ref. 157) is apparently a very optimistic limit. Is there tighter bound to accurately describe the fundamental limit of NFRHT? There have been some recent efforts to address this question[166,167,168]. Second, from the physical structure design side, can we design the physical structures to use the $\omega$–$k$ space more effectively and significantly go beyond the existing designs? A promising direction to addres this equstion is through inverse design of complex nanostructures[169] (Fig. 4i).

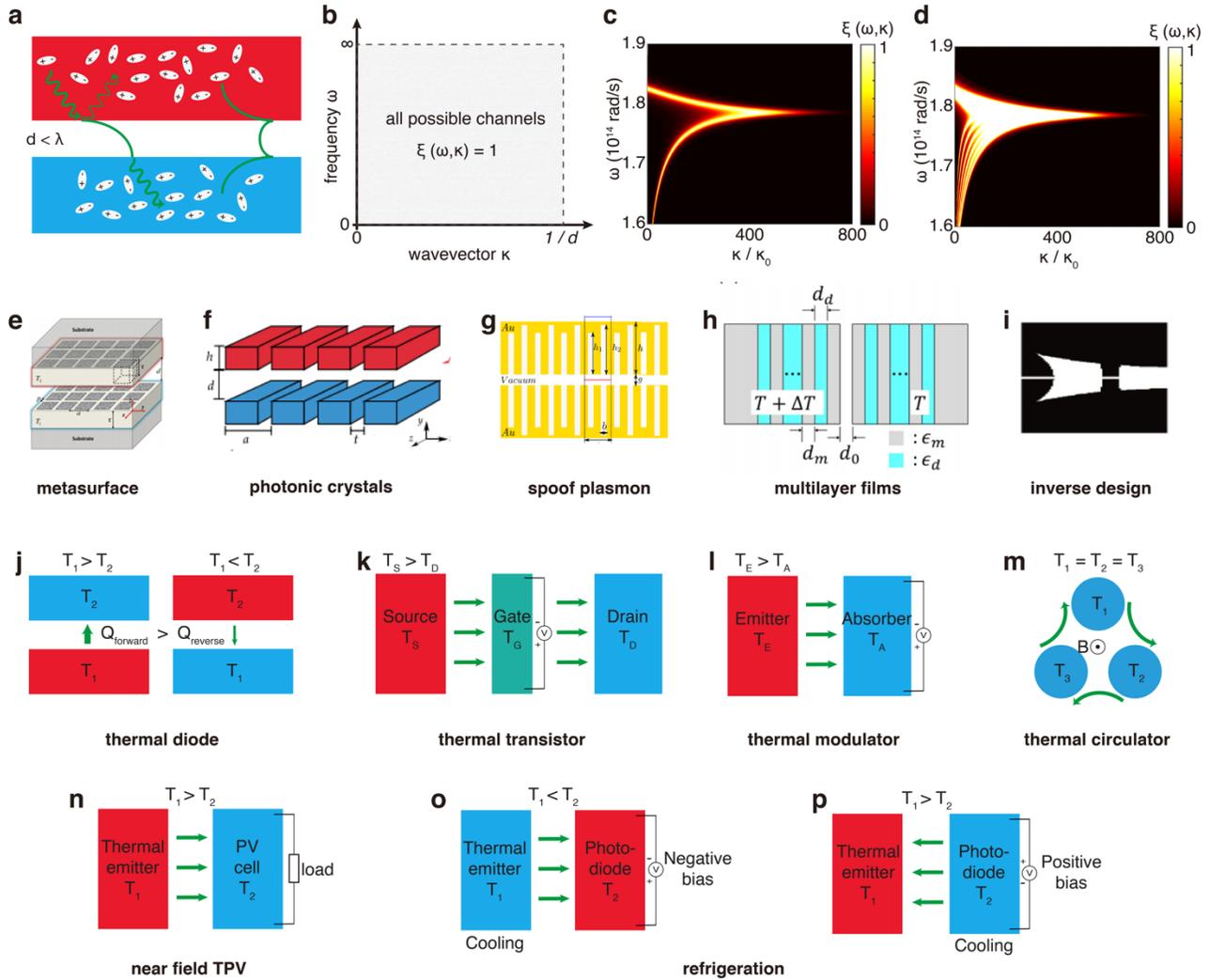

**Figure 4 | Near field thermal radiation manipulation**. **a |** Schematic of near field radiative heat transfer across a vacuum gap comparable to or smaller than the peak thermal wavelength. Evanescent waves and surface waves provide additional heat transfer channels and significantly enhance heat transfer. **b |** Theoretical limit of near field heat transfer with all possible channels over the $\omega$–$k$ space. **c |** A plot of photon transmission probability $\xi(\omega,k)$ between a pair of SiC with a gap distance of 10 nm. A standard surface mode enhances the heat transfer in the large $k$ regime but still only uses a small fraction of the $\omega$–$k$ space. **d |** Near field heat transfer enhancement with multiple modes constructed to fill up a larger region in the $\omega$–$k$ space. **e-i |** Designs for near field heat transfer enhancement including metasurface (**e**), photonic crystals (**f**), grooves (**g**), multilayer films (**h**), and inverse design structures (**i**). **j-p |** Near field heat transfer device concepts including thermal diode (**j**), thermal transistor (**k**), thermal modulator (**l**), thermal circulator (**m**), near field thermophotovoltaics (**n**), near field negative luminescent refrigeration (**o**), and near field positive luminescent refrigeration (**p**).

***Near field heat transfer manipulation.*** With the significant enhancement beyond far-field, NFRHT has become a promising approach for manipulating heat transfer at nanoscale. In analog to the key components in electronic circuits, here we review key components for near field heat manipulation including thermal diode, transistor, modulator, and circulator, with potential applications in nanoscale thermal management, and energy harvesting.

A thermal diode is a two-terminal device with the magnitude of thermal conductance depending on the direction of heat flow (Fig. 4j). To realize an NFRHT thermal diode, the key is to use materials or structures with different temperature dependency in their electromagnetic properties on the two terminals. For example, the different temperature dependence of two polytype SiC dielectric functions can lead to either alignment or misalignments of SPhP resonances across the vacuum as one changes the direction of the thermal bias, resulting in NFRHT rectification[170]. Phase change materials can also be used for this purposes[248,171,172].

In a thermal transistor (Fig. 4k), temperature bias on a gate is used to modulate the the heat flux between the source and the drain. Implementing a gate with phase change material, thermal switching, modulating, and amplifying can be achieved[173]. In addition to thermal transistor, heat transfer can be directly modulated with a two-terminal thermal modulator (Fig. 4l), by controlling the material or resonant properties of at least one terminal. For NFRHT modulator, a lot of concepts in dynamic control of far field thermal radiation can be directly brought into NFRHT, including the use of phase change materials and field effects (Section IV).

Breaking reciprocity can have important consequences in NFRHT, for example, to realize a thermal circulator (Fig. 4m), where a three or four-port device that can direct radiative heat flow entering any ports to the next port. Realizing thermal circulators can have important implications in energy harvesting[133,134]. It has also been shown that a nonreciprocal many-body system, consisting of magneto-optical nanoparticles, can support an equilibrium persistent directional heat current in the absence of temperature gradient[174]. The circulating nature of the persistent directional heat current is important for constructing a thermal circulator and is related to other novel heat transfer effects such as thermal hall effect[175].

NFRHT also has important implications in energy applications. For instance, although far field TPV system has been demonstrated[148], one of the major challenges is that its power density is inherently limited by the emitter temperature and the associated far-field heat transfer. This challenge can be addressed by the near-field effects[176]. A near field TPV device with significant enhancement in the power output relative to the far field was demonstrated[177]. Progress has also been made in enhancement of the power density of near-field TPV systems by exploiting both photon tunneling and electron tunneling[178,179].

The concept of photon chemical potential plays an important role in manipulating thermal radiation, as we discussed in the far field section. For example, a negatively biased semiconductor, i.e. with negative photon chemical potential, can have a lower apparent temperature than its thermodynamic temperature and therefore can absorb heat from and cool off a surrounding emitter at a lower thermodynamic temperature. This process is known as negative luminescent refrigeration (Fig. 4o). Recently, it has been theoretically suggested[180] and experimentally demonstrated[181] that this effect can be significantly enhanced in the near field regime, opening up possibilities for solid-state refrigeration. A closely related concept is positive luminescent refrigeration (Fig. 4p), where a positively biased semiconductor, can have an apparent temperature much larger than its thermodynamic temperature and can pump heat to a surrounding absorber at a higher thermodynamic temperature. The combination of near field heat transfer and positive luminescent refrigeration can potentially lead to a significantly larger cooling power and performance[182]. However, it remains a more significant challenge due to the stringent requirement on the semiconductor luminescent efficiency.

### IV. Active manipulation of heat transfer with multi-physical effects

The active influences of other physical fields on heat transfer can be categorized as two kinds. One introduces additional energy source to the system, while the other modulates the material thermal properties of the system. The thermoelectric effects are of the first kind that generates heating/cooling power with electric current. Thermal convection is another fundamental form of heat transfer, but it could be regarded as the first kind of multi-physical effects with thermal energy carried by moving matter. The second kind utilizes the sensitivity of thermal conductivity or emissivity to external fields.

***Thermoelectric effects.*** The study on thermoelectric (TE) effects is an active field with rich results[19,183]. Most of the efforts were made to achieve high-efficiency power generation utilizing the Seebeck effect (voltage $\Delta V$

$= -\alpha \Delta T$ induced by temperature difference $\Delta T$, where $\alpha$ is the Seebeck coefficient). The attention on the thermal part is limited to lattice thermal conductivity[184] and Peltier cooling[185].

The lattice thermal conductivity ($\kappa_l$) is mostly studied under the purpose of its reduction for a high material figure of merit $zT = \alpha^2 \sigma T/\kappa$[184,185]. The thermal conductivity ($\kappa$) consists of both the electric ($\kappa_e$) and lattice ($\kappa_l$) part: $\kappa = \kappa_e + \kappa_l$. For metals, $\kappa_e$ usually obeys the Wiedemann-Franz (WF) law: $\kappa_e = L\sigma T$, where $L$ is the Lorenz number[186]. Therefore, $\sigma$ and $\kappa$ are positively related, while a low $\kappa_l$ resolves the conflict between the demands of a large $\sigma$ and a small $\kappa$ for high $zT$. The positive relation between $\sigma$ and $\kappa$ is an important issue in the design of multi-physical functionalities that simultaneously manipulate heat and electric current. It is favourable when the desired thermal and electric effects are similar, such as cloaking[59] and camouflaging[60], because the same material can serve as a good conductor or insulator for both fields. However, much more efforts are required to design different or even opposite effects for the two fields, such as cloaking heat flux while focusing electric current[187,188]. Some materials[189,190] can potentially simplify the design, where WF law is violated upwards[189] or downwards[190].

The Joule heating and Peltier cooling effects are practical active approaches to manipulate heat with electric current. Joule heating happens inside the material due to the inelastic scattering of charge carriers $g = \boldsymbol{E}\cdot\boldsymbol{j} = \alpha\nabla T \cdot \boldsymbol{j} + j^2/\sigma$, where $g$ is heat generation in unit volume, $\boldsymbol{E}$ is the electric field, and $\boldsymbol{j}$ is the electric current density. Modifications may be required when the device length is comparable to the inelastic mean free path[191,192]. The term $\alpha\nabla T \cdot \boldsymbol{j}$ comes from the Seebeck effect, but it is counterbalanced by the heat flux carried by the electric current $\Pi\boldsymbol{j}$, where $\Pi$ is the Peltier coefficient. The Onsager reciprocity gives the second Thomson relation $\Pi = \alpha T$ when time-reversal symmetry is preserved[193]. Because of charge conservation ($\nabla \cdot \boldsymbol{j} = 0$), we have

$$\alpha \nabla T \cdot \boldsymbol{j} - \nabla \cdot (\Pi \boldsymbol{j}) = -T \nabla \alpha \cdot \boldsymbol{j} \tag{10}$$

which equals zero if the Seebeck coefficient is spatially uniform. Then Joule heating is $g = j^2/\sigma$ (Fig. 5a). When $\alpha$ is different for materials A and B, there is an additional heat generation at their interface that is the Peltier effect. The heat generated per unit area is $q_s = (\Pi_A - \Pi_B)\boldsymbol{j}\cdot\boldsymbol{n}$ where $\boldsymbol{n}$ is the unit vector perpendicular to the interface pointing from A to B (Fig. 5a). When electric current passes from A to B, it is a cooling effect for $\Pi_A < \Pi_B$. The performance of a Peltier cooler is positively related to $zT$ when used for refrigeration that extracts heat from the cold side[193] However, a high thermal conductivity is preferred for active cooling to facilitate heat dissipation from the hot side, where high $zT$ may not be favorable and some metals are suggested[194]. Joule heating is the governing mechanism of many heaters and most photothermal effects (*e.g.* the field of thermo-plasmonics[195] uses plasmonic resonance to convert light into Joule heat). It is also the major problem that throttles the speed and compactness of electronic devices. Combined with Joule heating, Peltier effect can be used in conductive thermal metamaterials to generate various temperature profiles[196], but they require temperature measurement and feedback. By applying on human body, adaptive thermal camouflage can be achieved using thermoelectric modules[197]. Peltier module can also become a thermal switch[198]. At nanoscale, thermoelectric rectifier and transistor have been theoretically proposed to regulate heat with charge current, based on a nonlinear three-terminal model[199].

Nonuniform Seebeck coefficient also comes from its temperature dependence $\alpha(T)$. When it is unnegligible, equation (10) is $-T(d\alpha/dT)\nabla T \cdot \boldsymbol{j} = -\tau \nabla T \cdot \boldsymbol{j}$, where $\tau = T(d\alpha/dT) = d\Pi/dT - \alpha$ is the Thomson coefficient. Such Thomson effect is considered to design TE modules with graded materials that can have much better optimized performance than the uniform case in theory[200]. The Righi-Leduc effect or thermal Hall effect can be induced by magnetic field or be found in ferromagnets to generate heat flux perpendicular to the applied temperature gradient[201]. In addition to electric current, there are also growing interests on the thermal effects induced by or dependent on spin current[202,203]. In insulators, the magnon Hall effect on spin current also has a thermal counterpart[204].

***Field effects.*** Externally applied electric, magnetic, or mechanical fields can induce entropy change and thereby generate or extract heat in some materials, which are called caloric effects[205]. They are often characterized by the isothermal entropy change $\Delta S$. For a strong effect that can be practically used, most studies are focused on materials that undergo phase transitions with drastic entropy changes induced by external fields, including magnetic materials, ferroelectrics, and shape-memory alloys. The maximum value of $|\Delta S|$ can be estimated across the phase transition. Good caloric materials have |$\Delta S_{max}$| around 10 J kg$^{-1}$ K$^{-1}$. Recently, plastic crystals as barocaloric materials[206] were reported to have |$\Delta S_{max}$| over 300 J kg$^{-1}$ K$^{-1}$. The caloric effects only work in the transient regime when the external field is temporally changed and are usually reversible when the change is restored, except for the anomalous electrocaloric effect[207]. Maintaining and utilizing them require repeatedly changing the working conditions with working fluids or deformation[208] for refrigeration or heat pumping[209].

External fields can also induce changes of material thermal properties (Fig. 5b). We already mentioned that temperature dependence and thermoelectric effects can be utilized to build thermal switches[67]. The field-induced modulation of thermal conductivity is another important method. It can be achieved with several different mechanisms. Firstly, field-induced phase transitions can be an effective way to introduce fast changes of the thermal conductivity and heat capacity. The metal-to-insulator transition (MIT) of vanadium dioxide ($VO_2$) is widely studied[210]. It can be triggered or suppressed by electric field, mechanical strain, hydrogenation, and photoexcitation, making it a convenient thermal switcher[211] and memory[212]. However, it has been reported that the total thermal conductivity $\kappa$ of bulk $VO_2$ is almost constant or even decreases[213] across MIT[190]. Fortunately, an increase of $\kappa$ following the WF law can be partially restored by introducing tungsten doping[190]. For thin films of $VO_2$, $\kappa$ can also increase[211] from ~4 W m$^{-1}$ K$^{-1}$ to ~6 W m$^{-1}$ K$^{-1}$. Another mechanism is the reconfiguration of domain walls in ferroelectric thin films by electric fields. At room temperature, the thermal conductivity of a lead zirconate titanate (PZT) film was reported to decrease upon applied gate voltage[214]. For two-dimensional materials, it has been proposed that external electric field can increase the phonon-electron scattering and thereby reduce the lattice thermal conductivity[215].

Electrochemical methods can achieve reversible modulation of thermal conductivity, as demonstrated in the thin films of $LiCoO_2$[216], graphite[217], black phosphorus[218], and $MoS_2$[219,220] under lithiation. The effects are often anisotropic. A nanoscale thermal transistor was realized based on the large-ratio and reversible thermal switching in $MoS_2$ thin films[220]. The mechanisms include lithian induced disorders, rattler modes, and strain. Electrochemically induced 'tri-state' phase transitions have been observed in $SrCoO_x$ (SCO), accompanied with thermal conductivity raise or decrease in a tenfold range[221]. The effects of oxygenation and hydrogenation on SCO include disorders as well as changes in lattice parameter, lattic structure, and ionic size. In this work, ion gels were used instead of liquid electrolyte to realize a semi-solid state device.

In addition to using electric field, many other active methods also uniquely influence thermal conduction. At cryogenic temperature, it has been reported that the lattice thermal conductivity of diamagnetic semiconductor InSb is reduced by external magnetic field[222], which is attributed to the enhanced phonon-phonon scattering via anharmonic interatomic bonds. A multiferroic thermal diode was proposed that is controllable by modifying the excitations of electric polarization and magnetization with electric or magnetic fields[223]. Light illumination can trigger a three-fold thermal conductivity switching in azobenzene polymers through a crystal to liquid transition[224]. The strain effects on the thermal conductivity of nanostructures have been widely studied[225,226]. A switchable thermal interface has been designed with a solid-liquid hybrid structure, where the combination of copper microposts and water demonstrates low thermal resistance under moderate loading pressure[227]. Under high pressure, semiconductor-metal transitions of Si and Si(Ge) can be triggered, leading to significant thermal conductivity enhancements[228]. Extreme pressure also modifies the phonon dispersion in bulk $MoS_2$ by enhancing the interlayer interaction, thereby increases the cross-plane thermal conductivity[229]. At nanoscale, the van der Waals interaction between boron nanoribbons enhances thermal conduction, and is tunable by wetting with various solutions[230]. It is worth noting that the aforementioned works mostly modulate the heat carrier populations or scattering rates, which are limited in a small spectral range. An interesting work proposed to modulate the displacement amplitudes of vibrations in topologically networked materials[231]. The idea has been demonstrated through hydration of squid ring teeth-based bio-polymers that changes the cross-linked structure, with a reversible fourfold switching of thermal conductivity.

The modulation of thermal radiation with external fields has made significant progresses in recent years, thanks to the rapid development of photonics. If the resonant modes of a photonic structure influence its absorption, its thermal emission can be modulated by electric fields, typically through the modified charge carrier density. The mechanism has been demonstrated for quantum wells[21,232], graphene plasmonic resonators[233], plasmonic metasurfaces[234], etc. The emissivity spectra of these materials are quickly responsive to gate voltage, enabling dynamically tunable and polarization dependent[234] thermal emissions. The idea was extended to the near-field radiative heat transfer between graphene layers[235]. As mentioned in Section IV, gate voltage also modifies the photon chemical potential[180,181,182]. Electrically induced mechanical deformation of elastomer membrane with an infrared-reflecting coating has been proposed for adaptive radiative camouflage[236], inspired by the color-changing capability of cephalopod skin. Another usage of electric fields is proposed to intercalate $Li^+$ into $Li_4Ti_5O_{12}$ (LTO)[237], which results in good tunability of the emissivity over broadband. Magnetic field can be used to break the detail balance or Kirchhoff's law (see Section III). It can also induce a large reduction of near-field heat transfer by suppressing surface modes and exciting hyperbolic modes[238]. Giant thermal magnetoresistance was predicted that enhances the near-field heat transfer in InSb-Ag nanoparticles by orthogonal magnetic fields, due to the spectral shift of localized surface waves[239]. Such an enhancement was also predicted for near-field heat transfer between graphene sheets at microwave frequency because of the field-induced magnetoplasmon zero modes[240].

Since static electric or magnetic fields are often uniformly applied, ultraviolet light has been proposed to locally modulate the thermal emissivity of an infrared metamaterial with photosensitive ZnO layer[241]. The modulation speed can be significantly increased by using visible pulsed laser, which induces nanosecond pulsed thermal emission from unpatterned silicon and gallium arsenide[242]. The effect of electromagnetic waves is similar to the gate voltage that modifies the charge carrier density. The photonic structures can also be deformed by mechanical stimulations. A microelectromechanical system (MEMS) can spatiotemporally reconfigure a metamaterial infrared emitter[243]. A textile made of bimorph fibers coated by carbon nanotubes has been designed for the thermal management of human body[244]. Its deformation and thereby emission spectrum are sensitive to the humidity change due to sweating. The thermal radiation of some material is tunable with chemical stimulus. Biomimetic infrared camouflage coatings have been made from a cephalopod protein, which is reconfigurable through the reversible swelling induced by acid vapor[245].

Temperature dependence is another useful mechanism. The MIT of $VO_2$ is accompanied with a quick drop of emissivity in the mid-infrared[246], as well as a decreased near-field heat transfer efficiency[247]. Similar ideas also apply to other phase-change materials such as $Ge_2Sb_2Te_5$[248,249]. These materials have been proposed to realize a plethora of new effects such as radiative thermal camouflage[248,250], super Stefan-Boltzmann relation[249], radiative thermal rectification[251,252], and zero-differential thermal emittance[249], as well as new devices such as radiative thermal diode[251], thermal rectifier[252] thermal homeostasis[253], and self-adaptive radiative cooling system[254].

***Convective effects.*** Convection is a unique and fundamental form of heat transfer. It is impossible to cover such a broad field here, so we will limit the scope of this part to works that focus on artificially designed structures or materials for the purpose of thermal conduction manipulation. Many devices like thermal exchanger, heat pipe, and thermosyphon are thus not discussed, because they directly utilize thermal convection. A common convective effect on thermal conduction is through the contact of fluid flow with solid, which is characterized by the dimensionless Biot number $Bi = hL_c/\kappa_s$, where $L_c$ is the characteristic length and $\kappa_s$ is the thermal conductivity of the solid body. The heat transfer coefficient $h$ gives the heat flux through the solid surface as $q_s = h(T_\infty - T)$, where $T_\infty$ is the temperature of the fluid far from the surface. The heat flux generally exists on solid surfaces exposed to air and could influence heat conduction, making the performance of some conductive thermal metamaterials inferior to the ideal case[255]. Fluid flow can also enhance heat transfer (Fig. 5c). For example, the effective thermal conductivity of nanofluids could be enhanced by local convective fields induced by Brownian motions[256]. A thermal meta-device was designed with a circulating fluid field[22]. The effective thermal conductivity $\kappa^{eff}$ of the fluid field is proved to be proportional to the square root of the Péclet number $Pe = \Omega R^2/D$, where $\Omega$ is the rotation speed, $R$ is the radius, and $D$ is the diffusivity. For more robust control, it was proposed to push $\kappa^{eff} \to \infty$ (much larger than solid materials) by high-speed forced convection.

A major effect of thermal convection is that thermal energy is carried into or out of the control volume at rate $-\nabla\cdot(\rho \mathbf{v} c_p T)$, where $\mathbf{v}$ is the velocity field, and $c_p$ is the specific heat at constant pressure[257]. Other factors that needed to be considered are viscous dissipation (usually small for laminar flows) and pressure work (negligible for incompressible fluids). Therefore, the heat transfer equation becomes the convection-diffusion equation:

$$\rho c_p \partial T / \partial t = \nabla \cdot (\kappa \nabla T) - \rho c_p \mathbf{v} \cdot \nabla T \tag{11}$$

The effects of the velocity field can be considered as a special kind of multi-physical effects on thermal conduction, characterized by the Péclet number Pe = $|v|L_c/D$. For constant uniform velocity fields, a variable change $r' = r - vt$ restores the diffusive equation, so the temperature field has a moving profile following the background, *i.e.* the advection (Fig. 5c). For conductive thermal metamaterials, an immediate question is whether the previous design principles are still applicable, but the answer is negative. The reason is that although equation (11) is form invarient[258], the Navier-Stokes equations (momentum equations for incompressible fluids) are not. However, transformation theory applies for thermal convection in porous media with small permeability[259], where the momentum equations are the Darcy's law which preserves form invarience[260].

Recently, a new perspective was proposed to consider equation (11) as a Schrödinger-like equation[261,262], where the convective term was identified as a Hermitian component introduced into a non-Hermitian Hamiltonian. This observation leads to the realization of anti-parity-time (APT) symmetry, a concept derived from parity-time symmetry[263]. One interesting property of an APT symmetric heat transfer system is that advection effect is suppressed in the symmetric phase. It could be understood as a cancellation of convective effects inside the field (Pe) with that brought by the adjacent field (Bi). In some situations, it is favourable that the heat transfer in moving matter could remain the same as in stationary media for us to predict and control it in the framework of conduction.

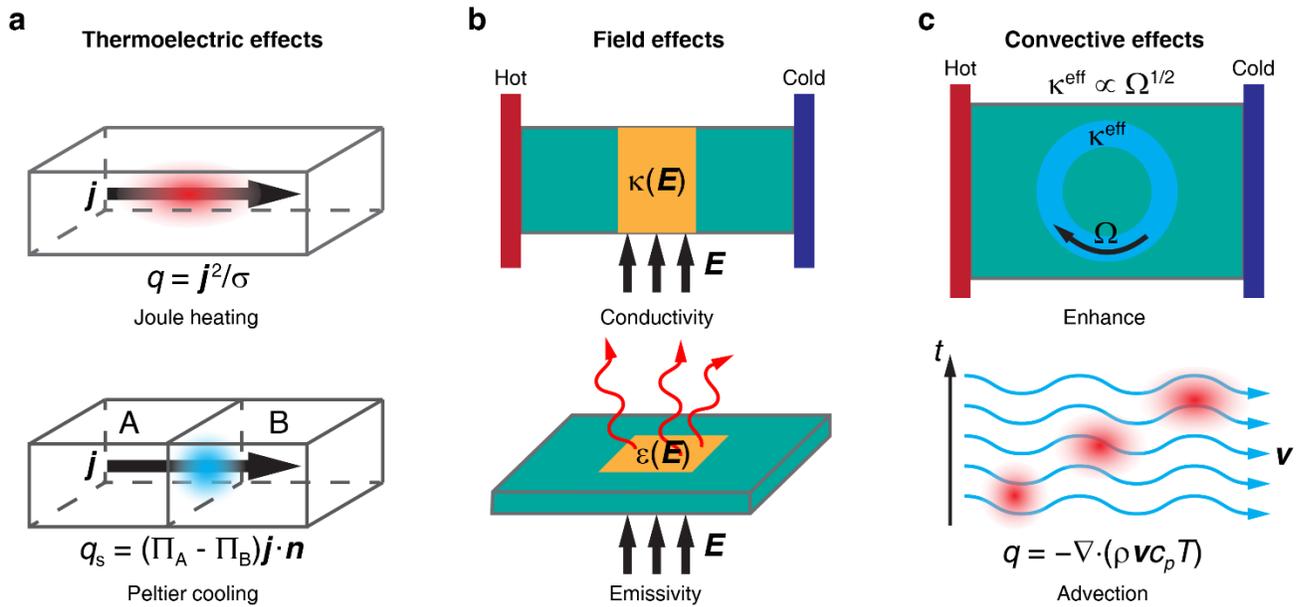

**Figure 5 | Multi-physical effects for heat transfer manipulation**. **a |** Electric current induces heat through the Joule heating effect or the Peltier effect. **b |** Electric (magnetic) fields can modify the thermal conductivity and emissivity of some materials. **c |** Moving matter (velocity fields) can enhance the thermal conduction in the contacting solid or bring the temperature field inside it in motion.

## V. Discussions and outlook

Heat transfer is a traditional topic with very long history. Nowadays' relevant research activities are mostly focusing on a specific form, either thermal conduction, radiation, or convection, at certain length scale (macroscale or nanoscale), and these activities are usually separated from other research fields.

In this review, we have tried our best to bring scattered works from different areas into the same umbrella. Some connections are worth noticing. There have been a few attempts to realize thermal cloaking at nanoscale based on the macroscopic designs[73,264]. Transformation theory has been used for the manipulation of radiative signals[37,39,54]. Convection has been used to enhance and cloak conductive heat[22]. The field-induced phase transitions are widely used to modify thermal conduction and radiation at all length scales. There are also theoretical designs of thermal metamaterials aiming at manipulating conductive and radiative heat together[265,266]. Despite of all these efforts, a big challenge is still to synthesize different approaches and results for more sophisticated and practical heat transfer control. It is difficult to fabricate large-scale nano-engineered phononic structures. Thermoelectric effects and caloric effects are seldomly considered as potential methods to modulate heat transfer beyond heating and refrigeration. Building such connections will not only enrich our knowledge, but also be helpful to many multiscale and multi-physical problems, such as heat dissipation in electronic devices and batteries, thermophotovoltaic energy harvesting, thermoelectric temperature regulation, etc.

Finally, we highlight some research interests that may bring new possibilities.

***Topological effects.*** The study on topological properties of material is an active research field. The phonon Hall effect[267] – a transverse heat flow induced by a magnetic field – was found to be the topological nature of phonons[268]. Inspired by electronic counterparts, phononic topological metamaterials that support topological protected one-way edge propagation[269] have been demonstrated. While most of these works focus on elastic wave, the unidirectional propagation of thermal phonons has not been achieved. It is still an open question if we can realize the thermal counterparts of topological insulator due to the small wavelength and coherent length of thermal phonons. Besides phonons, heat can also be carried by electrons, magnons, etc. Therefore, it is worthy to study the potential topological thermal effects induced by these particles/quasiparticles. Indeed, in electric systems, many topological insulators and semimetals are thermoelectric materials[270]. Also, chiral heat has been observed in quantum Hall systems[271]. The potential of topological states for improving thermoelectric performance is being actively investigated. In magnonic systems, magnon band topology and thermal magnon Hall effect has attracted intensive interests[272,273,274]. Topological heat transport was predicted in a two-dimensional bosonic Hofstadter lattice[275].

In the rapidly emerging field of topological photonics[276,277], new topological concepts are being exploited to design and control the behaviour of photons, leading to remarkable phenomena such as robust unidirectional propagation of light. These topological photonic ideas may open up new opportunities for photon-based heat flux manipulation. Along this direction, for example, there is an indication that the persistent heat current in many-body non-reciprocal heat transfer is connected to one-way edge states in these systems[278].

***Heat wave.*** A closely related subject is the wavelike behaviours of heat[279], which is the base of coherent scattering in NPC and local resonance in NPM (section II) and seems to be necessary for any topological effects according to the present theories. At macroscale, the diffusive Fourier's law indicates that any local perturbation of temperature will be instantaneously felt everywhere, which is against causality. Therefore, a rigorous governing equation must contain wavelike terms. The terms are often negligible because of the short relaxation time of the heat carriers. A promising direction to overcome the limit is the second sound[280] that was claimed to be observed in graphite between 85 and 125 K on the micrometre-length scale[281]. Numerical simulations even predicted the existence of second sound at room temperature in graphene and some other two-dimensional materials[282,283]. This makes it a promising mechanism to realize topological heat transfer[284]. At macroscale, it was proposed that the thermal convection in rotating structures could be regarded as a wavelike component[260].

***Quantum effects.*** As thermal devices reach the nano and atomic scale, the quantum counterparts of conventional thermal devices, such as quantum thermal diode[285], quantum thermal transistor[286] and quantum heat engine[287,288,289], have been proposed. An interesting question is whether we can make use of quantum effects, especially quantum coherence, tunnelling and quantum entanglement, to improve the performance of thermal devices and/or find new mechanisms to control heat. Quantum thermodynamics[290,291] might provide some insightful ideas. Using the trade-off between information and entropy, researchers experimentally demonstrated an on-chip information-powered refrigerator[289], and a spontaneous energy flow from the cold to the hot system in two quantum correlated spins-1/2[292]. It is verified that quantum coherence enables quantum heat engine to produce more power than its classical counterpart[293]. Although most investigations have shown that quantum effects can be useful in heat manipulation and conversion, these works focused on extremely microscopic systems (e.g. few spins system). It is interesting to investigate the scaling of quantum effect with system size, and furthermore, whether we can make quantum thermal devices in mesoscopic scale. To explore these subjects may require accurate control of many-body interactions and good scalability of the system. The fast development of quantum technologies provides some potential platforms, e.g. superconducting qubits, ion traps and optomechanics.

The fluctuational electrodynamics formalism is widely used to describe near-field heat transfer, which is driven by thermal fluctuation. The same formalism is also used to describe effects related to the quantum fluctuations of electromagnetic waves such as the Casimir effect, where momentum transfer and hence forces arise between neutral bodies. Thus, there is a strong connection in thermal fluctuation and quantum fluctuation effects. For example, with the presence of quantum fluctuations, phonon heat transfer between two objects separated by a vacuum gap can be achieved[294,295]. Exploration of quantum fluctuations could open up new opportunities for heat transfer manipulation in addition to conduction, convection, and radiation.

**Acknowledgements**

C.-W.Q. acknowledges the support from Ministry of Education, Singapore (Grant No.: R-263-000-E19-114). W.L. and S.F. acknowledge support by the U.S. Department of Energy (grant no. DE-FG02-07ER46426). W.L. acknowledges helpful discussions with Dr. Weiliang Jin and Lingling Fan.

**Author contributions**
These authors contributed equally: Y.L., W.L., T.H., and X.Z. All authors contributed to the writing, analysis, and deliberation. C.W.Q. supervised the project.

**Competing interests**
The authors declare no competing interests.